\documentclass[12pt]{iopart}
\usepackage{amsmath}
\usepackage{iopams}

\begin{document}

\maketitle

\title{Chiral Bargmann-Wigner Equations for Spin-1 Massive Fields}

\author{T. B Watson and Z. E. Musielak}
\address{Department of Physics, University of Texas 
at Arlington, Arlington, TX 76019, USA \\}
\ead{timothy.watson@mavs.uta.edu; zmusielak@uta.edu}

\begin{abstract}
The Bargman-Wigner equations are generalized to include chiral symmetry
based on the irreps of the Poincar\'{e} group, and the chirial Bargmann-Wigner 
equations are derived for spin-1 massive fields.  By specifying the chiral basis,
the chiral Bargmann-Wigner equations are reduced to the Proca-like equation, 
which is coupled by chirality to an auxiliary equation for spin-0 massive field.  
The coupling is a new phenomenon whose physical implications for the Higgs
field and dark matter are discussed. 
\end{abstract}

\section{Introduction} 
  
In the Bargmann-Wigner formalism [1], all possible relativistic equations are 
derived by using the unitary representations of the Poincar\'e group classified 
by Wigner [2,3].  A field of rest mass $m$ and spin $s \leq 1/2$ is described 
by a symmetric multispinor for which the Bargmann-Wigner (BW) equations 
are derived [1].  For special cases of $s = 1/2$, $s = 1$ and $s = 3/2$, the 
BW equations reduce to the Dirac [4], Proca [5] and Rarita-Schwinger [6]
equations, respectively; however, for $s = 2$, see [7].  The BW equations 
are coupled first-order partial differential equation with the original Dirac 
operator ($\hat D = i \gamma^{\mu} \partial_{\mu} - m$) acting on 
multispinor wavefunctions.

To generalize the Dirac equation, different modifications were made to the 
Dirac operator.  Physical reasons for these modifications range from attempts 
to unify leptons and quarks [8-10] and account for the three families of 
elementary particles [11-15], to including ad hoc a pseudoscalar mass [16] 
and extending the Dirac equation to distances comparable to the Planck length 
[17].  The main procedures giving such generalizations have been identified
as the Takuoka-Sen-Gupta [18-20], Weinberg-Tucker-Hammer [21-23] and 
Barut [11,12] formalisms [24].  Since each formalism leads to different forms 
of the Dirac equations and their opertaors $\hat D$, the corresponding 
modifications of the BW equations have also been accomplished [25].    

Recently, we derived the Dirac equation with chiral symmetry (DECS)  
by using the irreducible representations (irreps) of the Poincar\'{e} group 
$\mathcal {P}\ =\ SO (3,1) \otimes_s T (3+1)$, with $SO(3,1)$ being a 
non-invariant Lorentz group of rotations and boosts and $T(3+1)$ an 
invariant subgroup of spacetime translations, and the group including 
reversal of parity and time [3].  The derived DECS [26] can be written
\begin{equation}
(i \gamma^\mu \partial_\mu - m e^{-2 i \alpha \gamma^5})\psi = 0\ ,
\label{eq1}
\end{equation}
where $\alpha$ is the chiral angle and $\psi$ represents a four-component 
spinor that transforms as one of the irreps of $T (3+1) \in {\mathcal {P}}$
extended by parity, and each of its components satisfies the Klein-Gordon
equation [27,28].  We demonstrated that the chiral rotation of a massive 
field is equivalent to an alternative choice of chiral basis, and that the Dirac
equation is obtained by factorization of the Klein-Gordon equation if, and 
only if, a specific choice of chirial basis is selected [26].  

The nonstandard factorizations of the Klein-Gordon equation resulting from 
different choices of chiral bases redistribute the mass available to the field 
between the  left- and right-chiral components, which allows identifying 
the mass term in the DECS with pseudo-scalar mass.  The presence of this 
pseudo-scalar mass results in pseudo-scalar Higgs that can be used to explain 
smallness of neutrino masses and also properties dark matter particles [26].

In this Letter, we extend the DECS beyond spin-1/2 particles and generalize 
the BW equations to include chiral symmetry.  Our method to derive the 
chiral Bargmann-Wigner (CBW) equations is based on the irreps of the  
Poincar\'{e} group $\mathcal {P}$.  The derived CBW equations are 
valid for spin-1 massive fields.  By specifying the chiral basis, the CBW
equations reduce to the Proca-like equation [5] that is coupled to an 
auxiliary equation for spin-0 massive field is obtained.  The resulting 
coupling is caused by chirality and it is a new phenomenon, whose 
physical implications are discussed. 

\section{Bargmann-Wigner equations with chiral symmetry}

We may extend the usefulness of DECS beyond spin-1/2 particles by using multispinors
considered in the Bargmann-Wigner formalism [1].  We begin by observing that the at-rest 
solutions of Eq. (\ref{eq1}) may be written as
\begin{equation}
\psi^{(\pm)} = \omega^{(\pm)}(\alpha) e^{\mp i m t}\ ,
\label{eq2}
\end{equation}
where $ (+) \in \{ (1),(2) \} $ indicate positive energy solutions and $ (-) \in \{ (3),(4) \}$ 
indicate negative energies.  Then, these spinors take the form
\[
\omega^{(1)}(\alpha)=
\begin{bmatrix}
\cos{\alpha} \\
0 \\
i \sin{\alpha} \\
0\\
\end{bmatrix},\;
\omega^{(2)}(\alpha)=
\begin{bmatrix}
0 \\
\cos{\alpha} \\
0 \\
i \sin{\alpha}\\
\end{bmatrix},\;
\]
\begin{equation}
\omega^{(3)}(\alpha)=
\begin{bmatrix}
i \sin{\alpha} \\
0 \\
\cos{\alpha} \\
0\\
\end{bmatrix},\;
\omega^{(4)}(\alpha)=
\begin{bmatrix}
0 \\
i \sin{\alpha} \\
0 \\
\cos{\alpha} \\
\end{bmatrix}
\label{eq3}
\end{equation}
It is easily to verify that spinors $\omega^{(1)}$ and $\omega^{(3)}$ are $+\frac{1}{2}$ 
eigenstates of the spin projection operator $\hat{S}^{3}$ and $\omega^{(2)}$ and 
$\omega^{(4)}$ are similarly $-\frac{1}{2}$ eigenstates.   Moreover, all of these states 
have a definite propagation mass. 

We may now define the general set of positive energy multispinors of spin-1 from the tensor 
product of the spinors of spin-1/2 as follows
\begin{equation}
    \omega^{(1,1)}(\alpha,\beta) = \omega^{(1)}(\alpha)\otimes \omega^{(1)}(\beta)\ ,
\label{eq4}
\end{equation}
\begin{equation}
 \omega^{(1,2)}(\alpha,\beta)=\omega^{(2,1)}(\alpha,\beta) = \omega^{(1)}(\alpha) 
\otimes \omega^{(2)}(\beta) + \omega^{(2)}(\alpha) \otimes \omega^{(1)}(\beta)\ ,
\end{equation}
and
\begin{equation}
\omega^{(2,2)}(\alpha,\beta) = \omega^{(2)}(\alpha)\otimes \omega^{(2)}(\beta)\ ,
\label{eq6}
\end{equation}
where $\beta$ is the chiral angle assosciated with the chiral basis of the second bispinor of our 
representation. It must be noted that different chiral angles can be paired together in a single 
spinor as they are all valid eigenstates of momentum and spin as can be seen by observing 
these multispinors are eigenstates of the following spin operator (with indices included for clarity) 
\begin{equation}
\big(\hat{S}^3\big)^{\mu \nu}_{ \mu ' \nu '} = \big(\hat{S}^3 \big)^{\mu}_{ \mu'} 
\delta^{\nu}_{\nu'}+ \big(\hat{S}^3\big)^{\nu}_{ \nu'} \delta^{\mu}_{\mu'}\ ,
\label{eq7}
\end{equation}
and satisfy their respective at-rest DECS
\begin{equation}
(\gamma^{0} - e^{-2 i \alpha})^{\mu}_{\mu'} \omega^{(+,+)}_{\mu \nu}(\alpha,\beta) = 0\ ,
\label{eq7}
\end{equation}
and
\begin{equation}
(\gamma^{0} - e^{-2 i \beta})^{\nu}_{\nu'} \omega^{(+,+)}_{\mu \nu}(\alpha,\beta) = 0\ ,
\label{eq8}
\end{equation}
for $ (+,+) \in \{(1,1), (1,2), (2,2)\}$.  We may then move from the rest frame to an arbitrary 
momentum state by boosting each spinor
\begin{equation}
\omega^{(+,+)}_{\mu \nu}(\alpha,\beta;p^\mu) = \Lambda(\alpha,p^\mu)^{\mu}_{\mu'}
\Lambda(\beta,p^\mu)^{\nu}_{\nu'} \omega^{(+,+)}_{\mu \nu}(\alpha,\beta)\ .
\label{eq9}
\end{equation}
We note that the form of $\Lambda(\alpha,p^\mu)$ differs non-trivially from the form 
when $\alpha=0$. We omit a more detailed discussion of this form as, for our purposes,
the existence of such a transformation is sufficient.  We next define the symmetric and 
anti-symmetric positive energy multispinors and find
\begin{equation}
\Omega^{(+,+)} _{\mu \nu}(\alpha,\beta; p^\mu) =\frac{1}{2}
\big(\omega^{(+,+)} _{\mu \nu}(\alpha,\beta; p^\mu)
+\omega^{(+,+)} _{\mu \nu}(\beta,\alpha; p^\mu)\big)\ ,
\label{eq10}
\end{equation}
and
\begin{equation}
\widetilde{\Omega}^{(+,+)} _{\mu c}(\alpha,\beta; p^\mu) =
\frac{1}{2}\big(\omega^{(+,+)} _{\mu \nu}(\alpha,\beta; p^\mu)
-\omega^{(+,+)} _{\mu \nu}(\beta,\alpha; p^\mu)\big)\ ,
\label{eq11}
\end{equation}
such that the full symmetric and anti-symmetric positive energy solutions 
are given by
\begin{equation}
\psi^{(+)}_{\mu \nu} (\alpha,\beta, x^\mu) = \sum_{(+,+)} 
\int C_1^{(+,+)}(p^\mu)\Omega^{(+,+)}_{\mu \nu}(\alpha,\beta;
p^\mu) e^{-i p_\mu x^\mu} d^3 p\ ,
\label{eq12}
\end{equation}
and
\begin{equation}
\widetilde{\psi}^{(+)}_{\mu \nu} (\alpha,\beta, x^\mu) = 
\sum_{(+,+)} \int C_2^{(+,+)}(p^\mu)\widetilde{\Omega}^{(+,+)}_{\mu 
\nu}(\alpha,\beta;p^\mu) e^{-i p_\mu x^\mu} d^3 p\ ,
\label{eq13}
\end{equation}
where the sum is over $ (+,+) \in \{(1,1), (1,2), (2,2)\}$. 

Similar arguments for negative energy solutions lead to the construction of the negative energy multispinors. Notationally this is achieved via a simple index substitution $ (+,+) \rightarrow (-,-) \in \{(3,3), (3,4), (4,4)\}$ and flipping the sign of the exponential. With this complete set of states, we may identify the most general symmetric 
and antisymmetric multispinors
\begin{equation}
\Psi_{\mu \nu} (\alpha,\beta; x^\mu) = a_1 \psi^{(+)}_{\mu \nu} (\alpha,\beta, x^\mu)+ b_1 \psi^{(-)}_{\mu \nu} (\alpha,\beta, x^\mu) , \label{eq14}
\end{equation}
and
\begin{equation}
\widetilde{\Psi}_{\mu \nu} (\alpha,\beta; x^\mu) = a_2 \widetilde{\psi}^{(+)}_{\mu \nu} (\alpha,\beta, x^\mu)+ b_2 \widetilde{\psi}^{(-)}_{\mu \nu} (\alpha,\beta, x^\mu)
\label{eq15}
\end{equation}
that satisfy the following four coupled equations
\[
[i\gamma^\mu \partial_\mu - m \cos(\alpha-\beta) 
e^{-i(\alpha+\beta)\gamma^5}]^{\mu/\nu}_{\mu'/\nu'} 
\Psi _{\mu \nu}(\alpha,\beta; x^\mu) 
\]
\begin{equation}
\hskip0.25in = [-i m \sin(\alpha-\beta)\gamma^5 
e^{-i(\alpha+\beta)\gamma^5}]^{\mu/\nu}_{\mu'/\nu'} 
\widetilde{\Psi}_{\mu \nu}(\alpha,\beta;x^\mu)\ ,
\label{eq16}
\end{equation}
and
\[
[i\gamma^\mu \partial_\mu - m \cos(\alpha-\beta) 
e^{-i(\alpha+\beta)\gamma^5}]^{\mu/\nu}_{\mu'/\nu'} 
\widetilde{\Psi} _{\mu \nu}(\alpha,\beta; x^\mu) 
\]
\begin{equation}
\hskip0.25in = [-i m \sin(\alpha-\beta)\gamma^5 
e^{-i(\alpha+\beta)\gamma^5}]^{\mu/\nu}_{\mu'/\nu'} 
\Psi_{\mu \nu}(\alpha,\beta;x^\mu)\ ,
\label{eq17}
\end{equation}
where the summed indices have been combined for brevity. 

Isolating either multispinor yields
\begin{equation}
[\partial^\mu \partial_\mu + m^2]^{\mu}_{\mu '} 
\Psi _{\mu \nu}(\alpha,\beta; x^\mu) = 0\ ,
\label{eq18}
\end{equation}
and
\begin{equation}
[\partial^\mu \partial_\mu + m^2]^{\mu}_{\mu '} 
\widetilde{\Psi} _{\mu \nu}(\alpha,\beta; x^\mu) = 0\ ,
\label{eq19}
\end{equation}
which demonstrates that each element of our multispinors satisfy the Klein-Gordon equation.  
These are new BW equations with chiral symmetry (CBW equations) for spin-1 massive fields,
and they reduce to the BW equations when $\alpha = \beta$ in which case $\widetilde{\Psi}$ 
vanishes.  This shows that chiral symmetry requires additional equation for $\widetilde{\Psi}$.
The effects of this additional equation on the Proca equation [5] are now considered and 
discussed.

\section{New insights into the Proca Equation}

The CBW equations allows us to identify representations of spin-1 fields consistent 
with those degrees of freedom observed in the DECS.  Using [29], we relate these 
representations to those known in the particle physics. 
Let $\hat{C} = i \gamma^2 \gamma^0 $ and $\sigma^{\mu \nu} = 
\frac{i}{2} [\gamma^\mu,\gamma^\nu]$, then the Clifford basis is the 
following set of ten symmetric matrices $\{\gamma^\mu \hat{C}, 
\hat{\sigma}^{\mu \nu} \hat{C} \}$, and the six anti-symmetric matrices
$\{\gamma^\mu \gamma^5 \hat{C}, i \gamma^5 
\hat{C},\hat{C} \}$ that allow expanding the multi-spinors in this basis, and
obtian
\begin{equation}
\Psi = m A_\sigma \gamma^\sigma \hat{C} +  \frac{1}{2} F_{\sigma \tau} 
\hat{\sigma}^{\sigma \tau} \hat{C}\ ,
\label{eq20}
\end{equation}
and
\begin{equation}
\widetilde{\Psi} = \rho e^{-i \theta \gamma^5} \hat{C} + m B_\sigma 
\gamma^\sigma \gamma^5 \hat{C}\ ,
\label{eq21} 
\end{equation}
where $\rho$ is a scalar field, $\theta$ is a scalar parameter, $A^\mu$ 
and $B^\mu$ are vector fields, and $F^{\mu \nu}$ is an antisymmetric 
tensor. 

Writing the CBW equations in matrix form, we find the four linearly 
independent combinations
\[
i \partial_\mu \big(\gamma^\mu \Psi + \Psi(\gamma^\mu)^T\big) - m 
\cos{(\alpha-\beta)} \big\{\Psi, e^{-i(\alpha+\beta)\gamma^5}\big\} 
\]
\begin{equation}
\hskip1.0in - im \sin{(\alpha-\beta)} \big[\widetilde{\Psi},\gamma^5 
e^{-i(\alpha+\beta)\gamma^5}\big] = 0\ ,
\label{eq22} 
\end{equation}
\[
i \partial_\mu \big(\gamma^\mu \widetilde{\Psi} + \widetilde{\Psi}
(\gamma^\mu)^T\big) - m \cos{(\alpha-\beta)} \big\{\widetilde{\Psi}, 
e^{-i(\alpha+\beta)\gamma^5}\big\} 
\]
\begin{equation}
\hskip1.0in - im \sin{(\alpha-\beta)} \big[\Psi,
\gamma^5 e^{-i(\alpha+\beta)\gamma^5}\big] = 0\ ,
\label{eq23} 
\end{equation}
\[
i \partial_\mu \big(\gamma^\mu \Psi - \Psi(\gamma^\mu)^T\big) + m 
\cos{(\alpha-\beta)} \big[\Psi, e^{-i(\alpha+\beta)\gamma^5}\big] 
\]
\begin{equation}
\hskip1.0in + im 
\sin{(\alpha-\beta)} \big\{\widetilde{\Psi},\gamma^5 e^{-i(\alpha+\beta)
\gamma^5}\big\} = 0\ ,
\label{eq24} 
\end{equation}
and
\[
i \partial_\mu \big(\gamma^\mu \widetilde{\Psi} - \widetilde{\Psi}
(\gamma^\mu)^T\big) + m \cos{(\alpha-\beta)} \big[\widetilde{\Psi}, 
e^{-i(\alpha+\beta)\gamma^5}\big] 
\]
\begin{equation}
\hskip1.0in + im \sin{(\alpha-\beta)} \big\{\Psi,
\gamma^5 e^{-i(\alpha+\beta)\gamma^5}\big\} = 0\ .
\label{eq25} 
\end{equation}

Then, substituting the matrix expansions of the multispinors and exploiting the linear 
independence of the basis matrices, we obtain our constraints. Among these are the 
requirement that the fields $\rho$  and $B_\mu$ vanish and $\theta = 0$ unless we 
enforce $ \theta = \pi/2 - \alpha - \beta \ $. With this condition, we define the rotated fields 
\begin{equation}
{\begin{bmatrix}
        A'_\mu   \\
        i B'_\mu \\
    \end{bmatrix}}
    =
    {\begin{bmatrix}
        \cos{(\alpha -\beta)}    &   \sin{(\alpha -\beta)}                 \\
        -\sin{(\alpha -\beta)}   &   \cos{(\alpha -\beta)}               \\
    \end{bmatrix}}
{\begin{bmatrix}
        A_\mu   \\
        i B_\mu \\
    \end{bmatrix}}
\label{eq29} 
\end{equation}
and summarize the set of constraint equations as
\begin{equation}
\begin{aligned}
 F_{\sigma\tau} \cos{(\alpha+\beta)} +\frac{1}{2} \partial^\sigma 
F^{\mu\nu} \epsilon_{\mu \nu \sigma \tau} \sin{(\alpha+\beta)} - (
\partial_\sigma A'_\tau - \partial_\tau A'_\sigma ) &= 0\ ,
\\
\partial_\sigma B'_\tau - \partial_\tau B'_\sigma\ &= 0\ ,
\\
\partial^{\sigma}F_{\sigma \tau} + m^2 A'_\tau \cos{(\alpha+\beta)} &=0\ ,
\\
\frac{1}{2} \partial^{\sigma}F^{\mu \nu}\epsilon_{\mu \nu \sigma \tau} + 
m^2 A'_\tau \sin{(\alpha+\beta)} &=0\ ,
\\
\partial^{\mu}A'_{\mu}+i \rho \sin{(2\alpha - 2\beta)} &=0\ ,
\\
\partial^{\mu}B'_{\mu}+\rho \cos{(2\alpha - 2\beta)} &=0\ ,
\\
\partial_{\sigma}\rho - m^2 B'_{\sigma} &=0\ .
\end{aligned}
\end{equation}
Taking the divergence of the first of these constraints and eliminating explicit dependence on 
$F_{\mu\nu}$ and $B'_{\mu}$, we find the basic equations governing the fields 
reduce to three coulped equations of $A'_\mu$ and $\rho$:
\begin{equation}
\begin{aligned}
    \partial^\nu \big(\partial_\nu A'_\mu - \partial_\mu A'_\nu \big)
+m^2 A'_\mu &= 0\ , 
    \\
    \big[ \partial^\mu \partial_\mu + m^2 \cos(2\alpha-2\beta) \big] \rho  &= 0\ ,
    \\
    \partial^\mu A'_\mu + i \rho \sin{(2\alpha-2\beta)} &= 0\ .
\end{aligned}
\end{equation}
It is evident that the vanishing of $\rho$ is equivalent to the reduction of these equations 
to the Proca equation [5]. 

A more symmetric form of these equations is obtained by the definition of the constant 
$\kappa \equiv \sqrt{m^2 \cos{2 (\alpha - \beta)}-\mu^2}$ (where $\mu$ appears as 
a free parameter), and the scalar field $\varphi$ defined in terms of the divergence of 
the vector field $ \varphi \equiv \kappa^{-1} \partial^\mu A'_\mu$.  The constraint 
equations then reduce to  
\begin{equation}
\label{eq:finalConstraints}
\begin{aligned}
    \left[ \partial^\nu \partial_\nu +m^2\right] A'_\mu &= +\kappa \partial_\mu  \varphi\ ,
    \\
    \left[ \partial^\nu \partial_\nu +\mu^2\right] \varphi &= -\kappa \partial^\mu A'_\mu\ ,
\end{aligned}
\end{equation}
where the first equation is the Proca-like equation and the second is the auxiliary 
equation for $\varphi$.  Note that $m$ and $\mu$ are generally not equal and correspond 
to the masses of the vector and scalar fields respectively.  In the Lorentz gauge $\partial^\mu A'_\mu 
= 0$, the scalar wavefunction $\varphi = 0$ and the Proca-like equation becomes the 
Proca equation [5].  The coupling between the spin-1 and spin-0 massive fields is a 
new phenomena whose physical implications are now discussed.

\section{Physical implications}

The two main results of this Letter are the chiral Bargmann-Wigner (CBW) equations 
for spin-1 massive fields and the Proca-like equation with its required auxiliary equation 
for spin-0 massive fields.  There are several physical implications of these results. 

The degrees of freedom introduced by the choice of chiral basis allowed by Poincar\'e 
invariance admit an asymmetry to the defined representations of the considered spin-1 
massive fields described by the multispinors, thereby allowing for generalization of the 
BW equations to include chiral symmetry.  As a result two CBW equations are obtained, 
one for $\Psi$ and the other for $\widetilde{\Psi}$, and they are reduced to the original 
BW equations when $\widetilde{\Psi} = 0$, which is equaivalent to the case when chirality 
is neglected.  Thus, the main physical implication of the CBW equations is the presence 
of the additional field described by the multispinor $\widetilde{\Psi}$. 

To explore this additional field in detail, we specified the chirial basis in such a way that 
the CBW equations reduce to the Proca-like equation that is coupled to a spin-0 massive 
field.  We demonstrated that the asymmetry of the defined representations manifests 
itself physically as the coupled scalar and vector fields, with the total spin being consistent 
with our choice of representations.  By committing to a specific chiral basis one fixes the 
coupling between these fields and so restricts the system to the same total number 
of degrees of freedom as in the case when the chiral bases coincide.  The coupling 
is described by the coefficient $\kappa$ that depends on the chiral angles $\alpha$ 
and $\beta$ as well as on masses $m$ and $\mu$ of the scalar and vector fields,
respectively.

The coupling between the scalar and vector fields caused by the presence of chiral 
symmetry is a new phenomenon reported in this Letter.  While the fundamental 
spin-1 massive fields describing bosons $W^{\pm}$ and $Z^0$ are well-known 
in the Standard Model (SM) [29], the only fundamental scalar field in the model 
is the Higgs field [30,31], with a strong evidence for a elementary massive particle 
of spin-0 and positive parity [32].  Therefore, let us identify the scalar field coupled 
to the vector field is the Higgs field and explore the physical implications which 
follow from this supposition. Since 
our results show that the physical properties of the scalar field are such that its 
wavefunction $\varphi$ is proportional to the divergence of the vector field, with 
$\kappa$ being the proportionality coefficient representing chirality, this implies 
that the Higgs field must be related to the divergence of the vector wavefunction
and that chirality of spin-1 massive elementary particles place the dominant role
in this relationship.  To determine the validity of this statement further theoretical 
studies and potential experimental verifications are necessary but both out the 
scope of this Letter.

Another possibility is the existence of a scalar massive field representing the 
currently unexplained dark matter (DM) [33,34] and its physical properties that are 
likely to be described by a spin-0 massive elementary particle [35,36], whose
existence has not yet been verified experimentally [37,38].  The presence of 
such DM field and its possible coupling to spin-1 massive fields of ordinary 
matter (OM) through chirality would allow both OM and DM to be coupled.  
Moreover, as recently shown, in the nonrelativistic limit, DM may have both 
scalar [39] and vector [40] components that could be coupled by chirality.  
Further studies of these interesting phenomena are necessary and they will 
be described elsewhere.

Finally, let us point out that the form of the dependence of the coupling 
$\kappa$ on the chiral angles $\alpha$ and $\beta$ illustrates an important 
result of our derivation that we postulate holds in all physical systems, 
namely that \emph{only differences in chiral bases are experimentally 
observable}.

\section{Conclusions}

The original Bargman-Wigner equations are generalized by taking into account 
chiral symmetry for spin-1 massive fields.  The generalization is based on the 
irreps of the Poincar\'{e} group.  By specified the chiral bases, the derived chirial 
Bargmann-Wigner equations are reduced to the Proca-like equation, which is shown
to be coupled by chirality to an auxiliary equation for spin-0 massive field.  The 
physical implications of this coupling are discussed in the context of the scalar 
field to be either the Higgs field or a scalar massive field describing dark matter.  
In both cases, new and interesting results are likely to be obtained after more
detailed investigations are performed.


\section{References}
\begin{thebibliography}{}

\bibitem{1} V. Bergmann, E. Wigner, Proc. Natl. Acad. Sci. USA 34 (1948) 211
\bibitem{2} E.P. Wigner, Ann. Math. 40, 149 (1939)
\bibitem{3} Y.S. Kim and M.E. Noz, Theory and Applications of the 
                  Poincar\'e Group, Reidel, Dordrecht, 1986
\bibitem{4} P.A.M. Dirac, Proc. Royal Soc. London 117 (1928) 610
\bibitem{5} A, Proca, J. Physique et le Radiium 7 (1936) 347
\bibitem{6} W. Rarita, J. Schwinger, Phys. Rev. 60 (1941) 61
\bibitem{7} V.V. Dvoeglazov, Adv. App. Clifford Algebra 10 (2000) 7
\bibitem{8} I. Sogami, Prog. Theor. Phys. 66 (1981) 303
\bibitem{9} S.I. Kruglov, arXiv:hep-ph / 0507027v2 23 June 2006
\bibitem{10} E. Marsh, Y. Narita, Front. Phys. 3 (2015) Article 82
\bibitem{11} A.O. Barut, P. Cordero, G.C. Ghirardi, Phys. Rev. 182 (1969) 1844
\bibitem{12} A.O. Barut, Phys. Lett. 73B (1978) 310
\bibitem{13} W. Pfister, Nuovo Cimento A 108 (1995) 1427
\bibitem{14} S.I. Kruglov, Ann. Found. L. de Broglie 29 (2004) 1005
\bibitem{15} S.I. Kruglov, Phys. Let. B  718 (2012) 228
\bibitem{16} D. Leiter, G. Szamosi, Let. Nuovo Cim. 5 (1972) 814
\bibitem{17} K. Nozari, Chaos, Solitons \& Fractals 32 (2007) 302
\bibitem{18} Z. Tokuoka, Progr. Theor. Phys. 37 (1967) 581
\bibitem{19} N.D.S. Gupta, Nucl. Phys. B4 (1967) 147
\bibitem{20} A. Raspini, Fizika B5 (1996) 159
\bibitem{21} S. Weinberg, Phys. Rev. 133 (1964) B1318
\bibitem{22} S. Weinberg, Phys. Rev. 181 (1969) 1893
\bibitem{23} R.H. Tucker, C.L. Hammer, Phys. Rev. D3 (1971) 2448
\bibitem{24} V.V. Dvoeglazov, arXiv:hep-th/0208159v2  30 March 2003
\bibitem{25} V.V. Dvoeglazov, Int. J. Mod. Phys.: Cnference Series 2 
                   (2011) 121
\bibitem{26} T.B. Watson, Z.E. Musielak, Int. J. Mod. Phys. A 35 (2020) 2050189
\bibitem{27} O. Klein, Zeit. Phys. 37 (1926) 895
\bibitem{28} W. Gordon, Zeit. Phys. 40 (1926) 117
\bibitem{29} P. Renton, Electroweak Interactions, Cambridge Uni. Press, 
                    Cambridge, 1990
\bibitem{30} G. Aad, et al., ATLAS Collaboration, Phys. Lett. B 716 (2012) 1
\bibitem{31} S. Chatrchyan, et al., CMS Collaboration, Phys. Lett. B 716 (2012) 30
\bibitem{32} G. Aad, et al., ATLAS Collaboration, arXiv: 1307.1432v1 [hep-ex] 4 July 2013
\bibitem{33} K. Freeman, and G. McNamara, In Search of Dark Matter, 
                   Springer, Praxis, Chichester, 2006
\bibitem{34} J.A. Frieman, M.B. Turner, and D. Huterer, Ann. Rev. Astr. 
                   Astrophys. 46 (2008) 385
\bibitem{35} K. Sugita, Y. Okamoto, and M. Sekine, M., Int. J. Theor. Phys. 
                   47 (2008) 2875
\bibitem{36} R. Barbier, et al., Phys. Rep. 420 (2005) 1
\bibitem{37} M. Ackermann, et al., Phys. Rev. Let. 107 (2011) 241302
\bibitem{38} A. Ibarra, D. Tran, D. and C. Weniger, Int. J. Mod. Phys., 28 (2013)
                   1330040 (48pp)
\bibitem{39} Z.E. Musielak, Int. J. Mod. Phys. A, 28 (2021) 2150042 (12pp)
\bibitem{40} P. Adshead, and K.D. Lozanov, Phys. Rev. D, 103 (2021) 103501

\end{thebibliography}
\end{document}